# Experimental Observation of Modulation Instability and Optical Spatial Soliton Arrays in Soft Condensed Matter

P. J. Reece[a,*], E. M. Wright[b,a] and K. Dholakia[a,b]

[a] SUPA, School of Physics and Astronomy, University of St Andrews, St Andrews, Fife KY16 9SS, United Kingdom

[b] College of Optical Sciences, University of Arizona, Tucson, AZ85721

*Corresponding author:* pr20@st-andrews.ac.uk

**Abstract**

In this Letter we report observations of optically induced self-organization of colloidal arrays in the presence of un-patterned counter-propagating evanescent waves. The colloidal arrays formed along the laser propagation-axis are shown to be linked to the break-up of the incident field into optical spatial solitons, the lateral spacing of the arrays being related to modulation instability of the soft condensed matter system.

Optical spatial solitons (OSS) are spatially localized, non-diffracting modes that are supported in nonlinear optical media. They are the result of a balance between diffraction and self-phase modulation which results in self-focusing under intense illumination[1]. OSS have a number of distinct physical properties that differ from normal propagating modes in that they have a single, solitary beam-like profile and interact with other solitons in a particle like manner [2]. For a plane-wave incident field, small wavefront perturbations cause the optical field to break-up into periodic arrays of OSS [3,4] or more complex patterns [5], an effect known as modulational instability (MI) [6-8]. Both OSS and MI are generic properties of wave propagation goverened by the nonlinear Schrödinger equation. In addition to nonlinear optics, there are numerous examples in nature where such solitary waves and related phenomena are observed; particularly in pattern formation in granular systems [9] and complex fluids [10].

Recently Conti *et al.* suggested the possibility of OSS formation and MI in soft condensed matter systems, where local concentrations of constituent particles provide localised intensity dependent refractive index variations [11,12]. They developed a non-local theory which relates the Kerr response to a static structure factor, specific to the constituent materials. Colloidal dispersions have long been considered potential candidates for artificial Kerr materials. Ashkin and co-workers indicated the potential of bulk colloidal suspensions to act as artificial Kerr medium in a number of nonlinear optical experiments, including self-focusing, optical bistability and four-wave mixing [13-15]. The nonlinearity is an electrostrictive effect arising from optical gradient forces experienced by the dielectric particulates, which cause them to aggregate at regions of high intensity, thereby locally increasing the refractive index and leading to a self-focusing effect [16].

In this Letter we present an experimental system, based on an optical waveguide-geometry, to explore the nonlinear optical properties of large-scale colloidal aggregates. We show that the nonlinear behaviour of the system leads to observation of optically-induced self organization and particularly that colloidal arrays formed along the laser propagation-axis are linked to the break-up of the incident field into OSS, and that the lateral spacing of the arrays is related to modulation instability of the nonlinear coupled light-matter system. These results represent the first conclusive experimental realization of both MI and OSS arrays in colloidal suspensions [11,12]. We note Yashin *et al.* have indicated the formation of soliton-like beams bulk colloidal samples but stress that no conclusive demonstration of spatial solitons or any inference of MI was forthcoming in that study [16].

For the experimental work a prism-coupled resonant dielectric waveguide was used to generate counter-propagating (CP) waveguide modes with an evanescent wave (EW) component extending in to the colloidal solution (see figure 1). Transverse optical gradient forces due to the EW acting on particles in proximity to the supporting surface result in the accumulation of particles at the centre of the illuminated area, where the scattering forces due to the CP fields along the propagation (x) axis are balanced [17]. The dielectric waveguide constrains the field profile perpendicular to the waveguide to be the lowest order mode along the z-axis: this effectively eliminates one transverse dimension (z) leaving the propagation axis x and transverse dimension y. This particular design maximises the non-linear effects of the colloidal dispersion by using a waveguide design with a mode that has an enhanced field strength in the sample solution [18].

Resonant dielectric waveguides were fabricated by depositing an optical bi-layer coating of $SiO_2$ and $ZrO_2$ on to a high refractive index SF11 ($n_p$ = 1.754 @ 1064 nm) glass prism. In this configuration, the $ZrO_2$ layer formed the core of the waveguide, whilst the $SiO_2$ layer provided coupling to the prism and acted as the bottom cladding layer. The thickness and refractive index of the layers were chosen so that the waveguide supported a single TE mode (for $\lambda$=1064 nm) when an aqueous solution of colloidal particles is placed on the top the guiding layer. Colloidal preparations were made by diluting mono-dispersed 410 nm polymer colloids (Duke Scientific) in de-ionized water. A 100 μm thick circular sample chamber (1cm diameter) containing 20 μL of the solution was placed on the top of the prism in direct contact with coated surface.

Optical trapping was achieved using two CP waveguide modes excited through the prism coupler. In this geometry TE polarized light from a 1064 nm, Yb-doped fibre laser (IPG) was weakly focused onto the coupling layer / prism interface at an angle of approximately 60°; the spot size in at the sample surface in lateral direction was 100 μm. A second beam, coupled through the opposite prism face, was used to excite the opposing mode; the two beams where not temporally coherent. The relative power in each arm was also adjusted to achieve stable trapping at the centre of the trap; all powers quoted in the paper refer to the accumulative power in both arms. Observations of the colloid dynamics were monitored with a long working distance microscope objective (50x, NA=0.42 and 10x, NA=0.28) and a CCD camera.

Sub-micrometer colloids accumulated in the trap at high incident powers were observed to form regularly spaced linear arrays oriented along the direction of the incident EW wavevector with a lateral separation of several micrometers, whilst particles within the linear arrays were arranged in a disordered manner and exhibited strong thermal motion. The presence of large numbers of particles in the illuminated region was necessary to invoke the observed pattern formation. Video stills of the array formation are shown in figure 2 for a solution of 410 nm colloidal particles (concentration equivalent to 0.1% solid) accumulated in the counter-propagating cavity-enhanced evanescent wave trap (total incident power = 600 mW). Linear arrays, formed along the propagation direction of the evanescent waves, are regularly separated in the lateral direction with a periodicity of approximately 10 μm. We observe linear arrays under a range of experimental conditions including, different concentrations of colloids and different sized particles (200-600 μm diameter).

We associate and explain this new type of optically-induced self-organized array formation with the observation of arrays of OSS formed by the breakup of the waveguide mode through MI. (Due to the presence of waveguide losses and external pumping via the prism coupler we are strictly dealing with solitary waves as opposed to solitons, but we shall not dwell on the distinction here.) Qualitatively the observed ordering exhibits a number of properties that are suggestive of the presence of OSS. Firstly, a power threshold exists below which particles are accumulated but no ordering is observed, even over extended periods of illumination. Above this power threshold, stable arrays are formed with a characteristic spacing, which extend beyond the footprint of the illuminated area in the propagation direction. (With only one of the CP waveguide modes present the linear arrays still form but the colloids move along the direction of the mode under the action of the scattering force.) The spacing between the arrays in the lateral direction decreases with increasing power, until arrays become unstable and spontaneously evolve leading to merging and branching of arrays. A cross-section of the arrays, taken in the lateral direction and averaged over the propagation direction, clearly highlights the periodic nature of the linear array formation. From a simple analysis of the image the spacing between linear arrays is found to be 8.7 ± 0.6 μm. This characteristic spacing is similar for a range of different sample preparations, e.g. diluted samples (0.1%) and smaller sized colloids (200 nm): only the power threshold for observing the arrays changes significantly.

We now give a précis of the underlying theory for comparison with the experiment. Since array formation persists with only one of the CP waveguide modes present, and the CP modes are mutually incoherent, here we examine the unidirectional propagation case. Then for a prism coupled nonlinear planar waveguide, wave propagation along the x-axis may be described by the nonlinear Schrödinger equation (NLSE) [19,20],

$$\frac{\partial a_g}{\partial x} = \frac{i}{2k_0\beta_0}\frac{\partial^2 a_g}{\partial y^2} + ik_0\Delta\beta_0 \int_{-\infty}^{\infty} dy' G(y-y')\left|a_g(x,y')\right|^2 a_g + t_m a_{in} - \alpha_{eff} a_g \quad (1)$$

where $a_g(x,y)$ represents the amplitude of the guided wave, $k_0$ is the incident wavevector and $\beta_0$ is the effective-index of the guided wave. The first term on the right-hand-side of the NSLE describes beam diffraction in the transverse dimension (y-axis), the second term describes nonlinearities in the structure layers due to the colloids, with $G(y)$ a kernel that accounts for the nonlocality of the nonlinear response and $\Delta\beta_0$ the Kerr nonlinear coefficient for a local response with $G(y) = \delta(y)$, [12] the third term describes excitation of the guided wave by the incident field, $a_{in}(x,y)$, through the prism with transmission coefficient, $t_m$, and the last term describes losses due to both medium absorption and re-radiation, $\alpha_{eff}$, from the guided wave back into the prism. Detailed expressions for the above coefficients are given Liao et al.[19] and Stegeman et al. [20].

To understand the pattern forming properties of the nonlinear prism coupler we consider the stability properties of the high power plane-wave solution with respect to spatial modulations along the lateral dimension.[6-8] To proceed we consider a perturbed plane-wave solution of the form $a_g(x,y) = a_0 + a_+ \exp(2\pi i y/L) + a_- \exp(-2\pi i y/L)$, where we have added weak spatial side-bands to the strong plane-wave solution $a_0$ of Eq. 1, obtained by setting the spatial derivatives to zero. If the spatial side-bands grow they will create a spatial modulation of the intensity, and hence colloid density, of period L along the lateral or y-dimension. Substituting the perturbed plane-wave solution into the NLSE (1), and linearizing in the weak spatial side-bands, one obtains four-wave parametrically coupled equations for the side-bands. By demanding that the four-wave parametric interaction is phase matched we obtain the condition on the spatial period $L = \pi/(k_0\sqrt{\beta_0 \Delta\beta_0 S(0) |a_0|^2})$ of the MI, with S(q) the static structure factor which is the Fourier transform of the kernel G(y). The gain of the parametric interaction may be found by seeking a solution for the spatial side-bands of the form $a_\pm(x) = b_\pm \exp(gx)$, which yields for the parametric gain $g = k_0 \Delta\beta_0 |S(4\pi/L)||a_0|^2 - \alpha_{eff}$, so the power in the guided wave needs to exceed the threshold value $(|a_0|^2)_{th} = \alpha_{eff}/(k_0\Delta\beta_0|S(4\pi/L)|)$ for the growth of arrays to be possible. Thus, at threshold the lateral array spacing will be

$$L_{th} = \frac{\pi}{k_0\sqrt{\beta_0\Delta\beta_0 S(0)(|a_0|^2)_{th}}} = \frac{\pi}{\sqrt{\beta_0 k_0 \eta \alpha_{eff}}} \quad (2)$$

with $\eta = S(0)/S(4\pi/L)$.

From the theoretical analysis it is clear that the combined optical properties of the underlying dielectric waveguide and the colloidal suspension, and in particular the effective loss coefficient of the system, play an important role in determining the properties of the linear arrays. By experimentally determining the loss coefficient $\alpha_{eff}$ we may predict the lateral array spacing (2) at threshold that can be compared with the experiment for internal consistency. From a fit of the experimentally measured variation of the low power reflectivity of the prism coupled waveguide versus incident angle (not shown) we estimate the effective loss-coefficient as $\alpha_{eff}$ =110 cm$^{-1}$. Using Eq. (2) with $\eta$=1 we obtain a threshold periodicity $L_{th}$ =10 µm, in reasonable agreement with the experimental value. This implies that for our current experiment the lateral array spacing is sufficiently large that nonlocality is not a major player, since for a local response $G(y) = \delta(y)$ we have S(q)= const., and $\eta$=1.

From the model we also find the array spacing above threshold is power dependent. The relation $L = \pi/(k_0\sqrt{\beta_0\Delta\beta_0 S(0)|a_0|^2})$ relates the lateral array spacing to the guided wave power $|a_0|^2$. Then using the threshold guided wave power $(|a_0|^2)_{th} = \alpha_{eff}/(k_0\Delta\beta_0|S(4\pi/L)|)$, and the steady state solution $a_0$ of Eq. 1 obtained with the spatial derivatives set to zero, we find the relations

$$\frac{L}{L_{th}} = \frac{1}{\sqrt{\xi}} \ , \ \frac{(1+\eta^2\xi^2)}{(1+\eta^2)}\xi = \frac{P_{in}}{P_{th}}, \ \xi = \frac{|a_0|^2}{(|a_0|^2)_{th}} \qquad (3)$$

where $P_{in}$ is the input power, and $P_{th}$ the threshold input power for array formation which can be determined experimentally. Plotting $L/L_{th}$ versus $P_{in}/P_{th}$ parametrically as a function of $\xi$ according to Eqs. (3) we have a prediction of how the lateral array spacing varies with input power. Within the confines of our Kerr nonlinear model the lateral array spacing decreases monotonically with increasing input power beyond the threshold. Figure 4 shows a plot of average array spacing as a function of the input power along with the theoretical prediction obtained with η=1. The experimental data has been normalized with respect to the threshold array spacing and threshold input power. We see that there is good overall agreement between the theory and experiment, and in particular that the array spacing decreases with increasing power as expected in the array formation is due to MI.

It is important to note that the ideas of OSS and MI arise here because we have employed a continuum approximation to the colloidal system, that is, we have approximated the colloidal system as soft matter and ignored its granular structure. This importance of the granular nature of the nonlinear medium, particular near the limit of continuum approximation is highlighted by our observation that the lateral spacing between the arrays begins to decrease with increasing particle size and the emergence of more complicated pattern formation when colloidal particles begin to cluster.

We further contend that even for colloidal systems for which the granular nature is more important, the ideas of pattern formation are still of use for intuiting the underlying physics. In particular, continuum models have previously been of great utility in understanding the pattern forming properties of granular media [21]. MI in nonlinear systems may also lead to more complex pattern formation, such as two-dimensional square lattices, hexagonal and spiral type formation; this may be achieved by driving the system well above the threshold [21]. It is also known that interaction forces do exist between solitons, such as attractive forces between in-phase coherent OSS [2]; these forces may play a role in the ordering of higher order solitons. Furthermore by driving the system externally, by for example, modulation of the input beam, different types of localised waves, such as oscillons, may be observed [9] .

An interesting parallel may be drawn between the OSS/MI interpretation of optical force induced self-organization of colloidal dispersions and optical force mediated particle-particle interactions observed in optical binding; particularly in light of the recent reports of two-dimensional checkerboard-like array formation in counter-propagating evanescent wave trapping geometry, similar to the one described here [22]. Optical binding arise when particles locally perturb the incident field through refocusing and/or scattering, to create new stable trapping position for adjacent particles to occupy. Whilst longitudinal binding (i.e. binding along the propagation axis) has been well characterized in a number of trapping geometries [23,24], the origin of lateral binding is still not well established. A non-linear systems approach may provide further insights into the observed pattern formation for such lateral binding, although in this case the granularity of the system would become increasingly important.

In conclusion we have presented experimental evidence for the observation of modulational instability and formation of optical spatial solitons arrays in soft condensed matter. Our experiments performed on sub-micron colloidal dispersions offer the first conclusive demonstration in such media of these nonlinear phenomena and the observations are supported by theoretical calculations. The experimental system that we have developed here is an excellent platform for exploring more general properties of optically-induced nonlinear instability in colloidal systems. Finally the light matter interaction has also been noted for smaller numbers of colloid to organise into equilibrium positions due to the gradient and scattering forces: our work may also link to such lateral (or transverse) "optical binding" which remains a promising but not well understood method for optically controlled self assembly of particles [22,25].

This work, as part of the European Science Foundation EUROCORES programme NOMSAN, was supported by funds from the UK Engineering and Physical Sciences Research Council and the EC Sixth framework programme and the European Commission 6th framework programme— NEST ADVENTURE Activity—, through Project ATOM-3D (Contract No. 508952). EMW is funded by the Joint Services Optical Program (JSOP).

Figure Captions

Figure 1: shows a schematic diagram outlining the geometry of the experimental set-up and a light scattering image of spontaneous ordering of 410nm polymer colloids seen under typical experimental conditions. The top left diagram indicates the orientation of coupling prism; the propagation direction of the two counter-propagating modes are along the x-direction and the observed array formation is aligned along the x-direction with a periodic spacing in the y-direction (top right). The bottom diagram shows the structure of the nonlinear prism-coupled slab-waveguide used for the experiments. The colloidal suspension forms the top cladding layer and the waveguide parameters were chosen to achieve maximum extension of the mode into the top layer.

Figure 2: shows the temporal evolution of the OSS formation for incident powers above the threshold. Each panel is a CCD image of the illuminated region at different times after the light is coupled to the system. Bright field imaging also confirms the presence of ordered colloids in regions of high intensity.

Figure 3: The graph on the left shows in the period of the array spacing as a function of the input power above threshold. The period ($L$) is normalized with respect to the period at threshold ($L_{th}$) and the input coupling power ($P$) is normalized to the power at threshold ($P_{th}$). The solid line indicates the variation as predicted by the continuum model. The three panels on the right show the array formations for different illumination intensities. Below the threshold the OSS are not observed (top), even after extended periods of illumination; just above threshold stable arrays of OSS are formed.

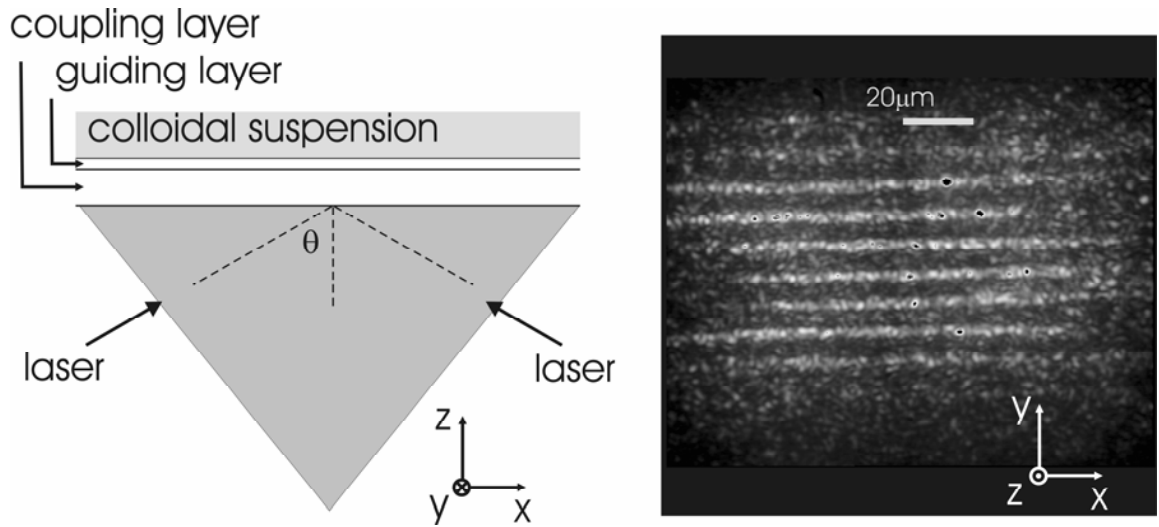

Figure 1. - P. J. Reece *et al.*

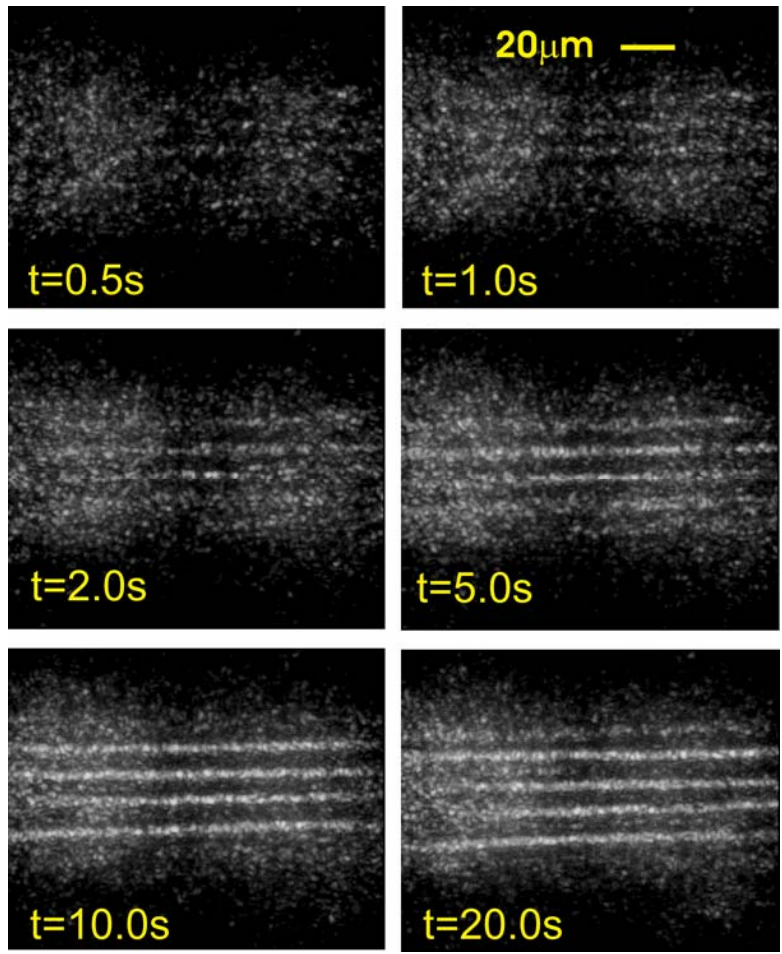

Figure 2 – P. J. Reece *et al.*

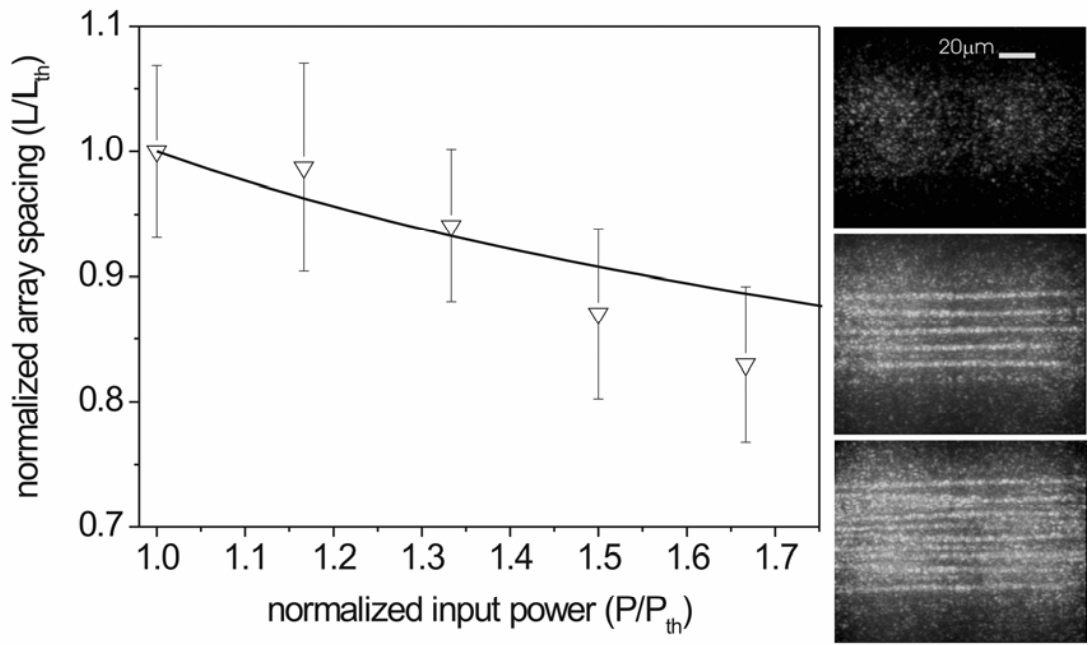

Figure 3 – P. J. Reece *et al.*